# Interface Dipole and Band Bending in Hybrid p-n Heterojunction MoS$_2$/GaN(0001)

Hugo Henck[1], Zeineb Ben Aziza[1], Olivia Zill[2], Debora Pierucci[3], Carl H. Naylor[4], Mathieu G. Silly[5], Noelle Gogneau[1], Fabrice Oehler[1], Stephane Collin[1], Julien Brault[6], Fausto Sirotti[5], François Bertran[5], Patrick Le Fèvre[5], Stéphane Berciaud[2], A.T Charlie Johnson[4], Emmanuel Lhuillier[7], Julien E. Rault[5] and Abdelkarim Ouerghi[1]

[1]Centre de Nanosciences et de Nanotechnologies, CNRS, Univ. Paris-Sud, Université Paris-Saclay, C2N – Marcoussis, 91460 Marcoussis, France
[2]Université de Strasbourg, CNRS, Institut de Physique et Chimie des Matériaux de Strasbourg (IPCMS), UMR 7504, F-67000 Strasbourg, France
[3]CELLS - ALBA Synchrotron Radiation Facility, Carrer de la Llum 2-26, 08290 Cerdanyola del Valles, Barcelona, Spain
[4]Department of Physics and Astronomy, University of Pennsylvania, 209S 33rd Street, Philadelphia, Pennsylvania 19104, USA
[5] Synchrotron-SOLEIL, Saint-Aubin, BP48, 91192 Gif sur Yvette Cedex, France
[6]Université Côte d'Azur, CNRS, CRHEA, 06560 Valbonne Sophia Antipolis, France
[7]Sorbonne Universités, UPMC Univ. Paris 06, CNRS-UMR 7588, Institut des NanoSciences de Paris, 4 place Jussieu, 75005 Paris, France

*Corresponding author, E-mail: mailto:abdelkarim.ouerghi@c2n.upsaclay.fr ,

**Abstract:**

Hybrid heterostructures based on bulk GaN and two-dimensional (2D) materials offer novel paths toward nanoelectronic devices with engineered features. Here, we study the electronic properties of a mixed-dimensional heterostructure composed of intrinsic n-doped MoS$_2$ flakes transferred on p-doped GaN(0001) layers. Based on angle-resolved photoemission spectroscopy (ARPES) and high resolution X-ray photoemission spectroscopy (HR-XPS), we investigate the electronic structure modification induced by the interlayer interactions in MoS$_2$/GaN heterostructure. In particular, a shift of the valence band with respect to the Fermi level for MoS$_2$/GaN heterostructure is observed; which is the signature of a charge transfer from the 2D monolayer MoS$_2$ to GaN. ARPES and HR-XPS revealed an interface dipole associated with local charge transfer from the GaN layer to the MoS$_2$ monolayer. Valence and conduction band offsets between MoS$_2$ and GaN are determined to be 0.77 and -0.51 eV, respectively. Based on the measured work functions and band bendings, we establish the formation of an interface dipole between GaN and MoS$_2$ of 0.2 eV.

## I. INTRODUCTION

Among the vast collection of two-dimensional (2D) materials, transition metal dichalcogenides (TMDs) have attracted considerable interest for their unique layer-dependent electronic and optical properties[1,2]. TMDs such as $MoS_2$, $MoSe_2$, $WS_2$ and $WSe_2$ have tunable bandgaps from indirect in their bulk form to direct in the monolayer limit, then opening up their wide range of potential applications in nano- (opto-) electronics. For example, $MoS_2$, one of the most studied TMDs, has been used in field effect transistors[3] with excellent on/off ratio and room temperature mobility and in photodetectors[4] with high responsivity and fast photoresponse. On the other hand, the p-n junction is an elementary block of optoelectronics and its demonstration using 2D TMDs is a mandatory step toward the integration of these materials in real devices[5–10]. Interestingly, the combination of 2D materials grown on conventional 3D semiconductor is gaining importance for the design of electronic devices, since it combines the advantages of both the established 3D semiconductors and the unique properties of 2D materials. An interesting combination can be obtained using bulk semiconducting GaN and 2D materials[11–15] in the so called mixed-dimensional heterostructures[16], due to the maturity of planar GaN technology with a broad range of devices spanning from light emitting diodes to high power electronics[17]. Hence, hybridation of GaN with 2D TMDs such as $MoS_2$ is of particular relevance to design novel hybrid heterostructures. Since theoretical studies on such heterostructures are particularly challenging, experimental research works are mandatory to uncover the 2D TMD/3D heterostructure interfacial and electronic properties and trigger further theoretical efforts. This paper is dedicated to get deeper insight on the electronic properties of $MoS_2$/GaN heterostructure as well as interlayer interaction (*i.e.* charge transfer) between the two building blocks.

The GaN substrate is suitable for opto-electronic applications. The use of 2D materials and GaN demonstrates examples of a 2D/3D combination matching the general requirements for the vertical heterojunction bipolar transistor. When considering the devices architecture, the interaction between the 2D layered film and the substrate is crucial. Similarly to 2D van der Waals (vdW) heterostructures[18,19], two key issues have to be considered : the strain effect caused by the lattice mismatch between both materials constituting the heterostructures, and the band offsets resulting from the junction formation. Only two works by Tangi *et. al.* focused on the interface between $GaN/MoS_2$[15] and $GaN/WSe_2$[14]. The authors have grown undoped GaN on 2D materials and performed micro-Raman and X-ray photoemission spectroscopy (XPS) to investigate the properties of band alignment in these heterostructures. However for the optoelectronic applications, a n- or p-doped GaN is required. Ruzmetov D. *et al.*[11] have grown $MoS_2$ on n-type GaN/sapphire. Using conductive AFM (CAFM) they showed that the $MoS_2$/GaN heterostructure electrically conduct in the out-of-plane direction and across the van der Waals gap, forming a promising platform for vertical 2D/3D semiconducting devices. Moreover, Lee II E.W. *et al.*[13] realized a p-$MoS_2$/n-GaN heterojunction diodes. No Fermi level pinning was present at the interface and current-voltage measurement of the diodes exhibited excellent rectification. Besides, the influence of the stacking order can modify the electronic properties at the interface. In the meantime, to our knowledge, no work in the literature was performed using ARPES to study the electronic structure such as charge transfer, interface dipole and band bending at n-doped $MoS_2$ on p-doped GaN interface.

Based on these considerations, the impact of p-doped GaN as a substrate should not be overlooked. The investigation of electronic properties of MoS$_2$ combined with GaN becomes of fundamental importance. Therefore, based on Raman spectroscopy, we assess the strain sustained by the MoS$_2$ flakes, upon transfer on top of GaN. Next, by using angle resolved photoemission spectroscopy (ARPES), we show a significant charge transfer between MoS$_2$ monolayer and GaN(0001) layer. ARPES measurements showed that the GaN valence band maximum (VBM) shifts of about 300 meV towards the Fermi level compared to the VBM of pristine GaN(0001); implying electron transfer from the GaN layer to MoS$_2$. Thus, we expect that this experimental study, which offers a better understanding of these heterostructures, will provide sound guidelines towards real industrial applications and complement the recently introduced 2D/2D approaches[20,21] for which device growth and processing remain quite challenging to scale up.

## II. METHODS:

The 250 nm thick p-doped GaN was grown by plasma-assisted molecular beam epitaxy (MBE) on a SiC(0001) substrate. The growth was performed at 730 °C under Ga-rich conditions to favour the 2D growth following the Frank Van der Merve growth method. During the growth, the Mg cell was kept at 375 °C in order to induce a p-type doping of GaN layer[22]. Large scale MoS$_2$ monolayer flakes (≈20 to ≈100 μm) have been grown by Chemical Vapor Deposition (CVD) on oxidized silicon substrate (see methods and ref[23]). The MoS$_2$ flakes transferred onto the GaN retain their triangular shapes with unchanged lateral sizes. Before any measurement, the MoS$_2$ sample was annealed at 300 °C for 30 min in ultra-high vacuum (P ≈ 10$^{-10}$ mbar), in order to remove the residual surface contaminations induced by the wet transfer. The Raman and PL measurements were conducted using a commercial confocal Renishaw micro-Raman microscope with a 532 nm laser in ambient conditions of pressure and temperature. The excitation laser (wavelength 532 nm) was focused onto the samples with a spot diameter of ~1 μm and incident power of ~3 mW. The integration time was optimized in order to obtain an acceptable signal-to-noise ratio. PL measurements were performed on the same microscope with a 100× objective and a Si detector (detection range up to ~ 2.2 eV). XPS experiments were carried out on the TEMPO beamline (SOLEIL French synchrotron facility) at room temperature using a photon energy of 340 eV. The photon source was a HU80 Apple II undulator set to deliver linearly polarized light. The photon energy was selected using a high-resolution plane grating monochromator, with a resolving power E/ΔE that can reach 15,000 on the whole energy range (45 - 1500 eV). During the XPS measurements, the photoelectrons were detected at 0° from the sample surface normal $\vec{n}$ and at 46° from the polarization vector $\vec{E}$. The spot size was about 100 × 80 (H×V) μm$^2$. A Shirley background was subtracted in all core level spectra. The Mo 3d spectra were fitted by sums of Voigt curves, i.e, the convolution of a Gaussian (of full-width at half-maximum GW) by a Lorentzian (of full-width at half-maximum LW). The LW was fixed at 90 meV. The Ga 3d was fitted with a Voigt curve. The ARPES measurements were conducted at the CASSIOPEE beamline of Synchrotron SOLEIL. We used horizontal linearly polarized photons of 50 eV and a hemispherical electron analyzer with vertical slits to allow band mapping. The total angle and energy resolutions were 0.25° and 25 meV. The mean diameter of the incident photon beam was smaller than 50 μm. All ARPES experiments were done at room temperature. The Fermi level reference was taken at the leading edge of a clean metal (i.e. molybdenum clamps) surface in electrical contact with the sample. The CL measurements were performed with an Attolight Chronos cathodoluminescence microscope operating at room temperature with 2 kV acceleration voltages. An IHR320 spectrometer (Jobin Yvon) coupled to a Newton CCD 920 camera (Andor/Oxford Instrument) was used to acquire the CL spectra. The final hypermap is a square matrix of 64x64 pixels, spanning about 30 microns wide, each of them containing a CL spectrum acquired in 1s. The simulations used to evaluate the CL excitation volume at 2 kV were performed using the CASINO v2.48 software (2D version), using a density of 6.15 g cm$^{-3}$ for the GaN material.

## III. RESULTS AND DISCUSSIONS

A 250 nm thick Mg-doped GaN(0001) (p-doped) grown by plasma-assisted molecular beam epitaxy (MBE) on SiC(0001) was used as a substrate. A few hundred micrometer wide monolayer $MoS_2$ flakes, grown by chemical vapor deposition (CVD) on $SiO_2$ substrate were transferred on the GaN layer by PMMA assisted technique[18] to build a $MoS_2$/GaN heterostructure as shown in Figure 1(a). An annealing process at T= 300 °C for 30 min was then used to clean the surface and the interface of our 2D/3D heterostructure. Owing to the large optical absorption of monolayer $MoS_2$, it is rather simple to identify $MoS_2$ flakes on the GaN surface as shown in the optical image in Figure 1(b). Hence, we can verify that the geometry and the sizes of the $MoS_2$ flakes are preserved during both transfer and annealing processes.

In Figure 1(c), we show the micro-Raman spectra, in the wavenumber range between 360 and 440 cm$^{-1}$, obtained for the $MoS_2$ transferred on GaN (black line)[1]. We can identify the two one-phonon Raman-active modes of monolayer $MoS_2$, namely the in-plane (E') and out-of-plane ($A'_1$) modes.[24,25] The frequency difference between the frequencies of the $A'_1$ and E' mode-features is ≈19-20 cm$^{-1}$, a value that is typical from pristine monolayer $MoS_2$ (see Fig. 1d).[18,26] The Raman intensity maps of the E' and $A'_1$ modes and the corresponding Raman frequency maps are shown in the supporting information Figure S1. The uniform intensity of both Raman modes illustrates the high quality and the absence of defects in our $MoS_2$ monolayers. The E' and $A'_1$ mode-frequencies display only minute spatial inhomogeneity of ≈ ± 1 cm$^{-1}$ over a given $MoS_2$ single-domain (see Figure 1(d)). These results indicate that inhomogeneous strain due to the $MoS_2$ transfer process can be neglected. Figure 1(e) shows the photoluminescence (PL) spectrum for $MoS_2$/GaN measured at room temperature. On the PL spectrum, we identify the well-known A and B excitons located near 1.84 and 2 eV, respectively.[27] The A exciton energy is assigned to the optical band gap of $MoS_2$ on GaN, which is similar to the value found for van der Waals heterostructures such as $MoS_2$/graphene.[28]
X-ray photoemission spectroscopy (XPS) and ARPES measurements were carried out for pristine GaN and $MoS_2$/GaN(0001) sample not only to investigate the atomic composition and the chemical bonding environment of the interface of our samples, but also to uncover the interface-based electronic properties of this heterojunction (band bending, work function, and dipole).

Figure 2(a) shows the XPS spectra of Ga-3d for $MoS_2$/GaN(0001) and the pristine GaN. In the two cases, only one peak is present corresponding to the Ga-N bonds. No oxidation was observed since the Ga-3d spectrum did not show any corresponding peak (expected at 1-1.2 eV higher binding energy (BE) with respect to the Ga-N peak[29,30]). After the MoS$_2$ transfer on GaN, the Ga-3d peak shifts towards lower binding energy (about 200 meV). This shift at lower BE indicates a variation of the band bending, result of a charge redistribution at the MoS$_2$/GaN interface. This effect will be duly discussed in next sections.

The Mo-3d spectrum (in Figure 2(b)) contains one main doublet component at binding energy (BE) Mo 3d$_{5/2}$ = 229.7 eV (3d$_{5/2}$:3d$_{3/2}$ ratio of 0.66 and a spin-orbit splitting of 3.10 eV[31] ) related to a Mo$^{4+}$ in a sulphur environment[32]. At lower BE (~ -0.52 eV) with respect to this main doublet peaks a small component is present (highlighted in green) which is the signature of a defective/sub-stoichiometric MoS$_2$ with sulfur vacancies (Sv)[32,33]. The weight of this component (between 10-15% of the whole Mo 3d spectrum) is not representative of a single MoS$_2$ flake due to the large X-ray spot size (~100 × 80 (H×V) µm$^2$), but it gives an information on the percentage of defective MoS$_2$ in the explored area. The shoulder at BE = 226.5 eV represents the sulphur 2s peak. These BE values for the Mo 3d indicate an intrinsic n-type doping of the MoS$_2$ flakes[34]. No other components are present on the Mo 3d spectrum related to nitrogen, oxygen or carbon bonds[35–37] indicating that no contaminations are present on the sample and confirming the high quality of the interface of this hybrid heterostructure. Moreover, there is no signature of any chemical state associated with Mo or S in the Ga 3d core level spectrum which is a clear evidence of a van der Waals interaction between MoS$_2$ and GaN. This makes the heterointerface atomically abrupt without any inter-diffusion.

In order to investigate the band alignment and the electronic properties of the hybrid MoS$_2$/GaN heterostructure, we performed band structure measurements by angle-resolved photoemission spectroscopy (ARPES) at Cassiopée beamline of Synchrotron Soleil. The photoelectron intensity as a function of energy and k-momentum of pristine GaN(0001) and hybrid MoS$_2$/GaN(0001) are presented in Figure 3(a) and (b) respectively. The respective second-derivative spectra are provided in Figure 3(c) and (d) to improve the visibility of the band structure. The valence band structure of GaN is shown in Figure 3(a) and (c). From Figure 3(b) and (d), we notice the appearance of a new top-most band at around -1.5 eV, which is the signature of MoS$_2$ valence band. This confirms the high quality of the transferred MoS$_2$ within the hybrid heterostructure. We also can notice that the topmost band of GaN is upshifted upon the MoS$_2$ transfer.

In Figure 4 (a) is shown a vertical section at k$_{//}$= 0 Å$^{-1}$ of the band structure of the MoS$_2$/GaN heterostructure and the pristine GaN. From the intersection of the linear extrapolation of the leading edge of the valence band spectrum with the baseline, we can locate the position of the valence band maximum (VBM) for the GaN in the heterostructure with respect to the pristine one. The relative VBM positions moved from 2.59 to 2.27 ± 0.05 eV for the GaN(0001) layer and the MoS$_2$/GaN heterostructure, respectively. This value is in agreement with the observed bandshift in ARPES measurements. This implies that valence band maximum (VBM) is getting closer to the Fermi level (located at 0 eV), reducing the band bending (V$_{BB}$) at the interface of about $\Delta V_{BB} = 0.32$ eV. The valence band for the pristine GaN, was also measured with a photon energy hv =1300 eV (Figure S2). Using this photon energy, a probing depth of about 10 nm is reached, which is reasonably larger than the depletion region at the GaN interface. Then using the same procedure used in Figure 4, the distance of the valence band to the Fermi level E$_v$ = 0.7 ± 0.05 eV in the bulk (*i.e.* in a flat band condition) was obtained. Considering that at the surface the VBM = 2.27 eV, a downward band bending of about 1.57 eV is present at the GaN(0001) surface. This band bending corresponds to an accumulation of positive charge at the GaN surface, compensated by an opposite negative charge inside the semiconductor (*i.e.* depletion layer). When the MoS$_2$/GaN heterostructure is formed, this band bending is reduced. This effect is the result of electron transfer (interface dipole formation) from GaN(0001) in favor of MoS$_2$. From Figure 4(a) we are also able to infer more precisely the VBM for the MoS$_2$, VBM = 1.5 ± 0.05 eV, which implies a valence band offset between MoS$_2$ and GaN (ΔE$_v$) of about 0.77 eV. To gain insight into the electronic properties of the MoS$_2$/GaN interface, the work function of the heterostructure was compared to the work function of the pristine GaN *via* the measurement of the secondary electron cut off (Figure 4(b)). We found out a work function of ϕ=5.23 ± 0.05 eV for pristine GaN and ϕ=5.35 ± 0.05 eV for the hybrid MoS$_2$/GaN heterostructure. Based on literature about probing quasiparticle band structure by STM/STS[21], the band gap of MoS$_2$ is about 2.15 eV. This MoS$_2$ electronic band gap is larger than its optical band gap determined previously by PL spectroscopy (~1.84 eV, see Figure 1(e)) considering the large exciton binding energy in atomically thin TMDs[38]. The Cathodo-luminescence (CL) experiments described in Figure S3 were performed to further probe the MoS$_2$/GaN interface. From Figure 4(c), we determine a value of 3.41 ± 0.01 eV for the GaN optical gap at room temperature, consistent with the reported value of wurtzite GaN[39]. Considering the exciton binding energy in GaN of about 0.02 eV[29] we deduce a GaN excitonic gap of about 3.43 eV± 0.01 eV. Thus, with the known values of the band gaps (MoS$_2$ and GaN) the conduction band discontinuity ΔE$_C$ is calculated from: ΔE$_C$ = ΔE$_V$ − (E$_{MoS2}$ − E$_{GaN}$) where E$_{MoS2}$ and E$_{GaN}$ are the bandgap energies of MoS$_2$ and GaN, respectively; we obtain ΔE$_C$ = -0.51 eV with type II band alignment at n-MoS$_2$/p-GaN heterojunction. This conduction band offset is close to the recently reported value (0.56 eV) for intrinsic epitaxial GaN/MoS$_2$,[15] with an inverted band position with respect to our work ( *i.e.* the GaN CBM is at higher binding energy with respect to the MoS$_2$ CBM). A different band alignment was obtained in the case of n-doped GaN/p-doped MoS$_2$ [13] where a conduction band offset of 0.23 eV was measured. These results suggest the possibility of tuning the relative band alignment in the 2D/3D heterostructure and then the potential barrier height at the junction by varying the doping of the MoS$_2$ and the GaN layers. By combining all the photoemission studies an interface electronic structure diagram is derived (Figure 5). These findings are in agreement with what was observed for WS$_2$/p-doped GaN[40] where an efficient charge transfer at the 2D-3D heterointerface was observed. In particular, Kummel *et al.* have shown that for this 2D/3D heterostructure the efficiency of the charge transfer across the heterointerface is influenced by the momentum mismatch of the VBM in the two semiconductors. In particular, they underlined that the charge transfer process is more efficient when the excitation in the k space is near the Γ point of the 2D semiconductor, where a transfer to the 3D substrate is possible without a momentum change. At variance with metal, the work function of a semiconductor is not an intrinsic property, simply because the position of the Fermi level in the gap at the surface depends on the doping of the substrate which determines the amount of band bending. Moreover, when we form the MoS$_2$/GaN heterostructure a surface dipole (Δϕ$_{Dip}$) could be formed at the interface. This effect is described by the measured variation of work function of the system (Δϕ = 0.12 eV). We have to take into account that in the case of a semiconductor the variation of band bending at the surface (ΔV$_{BB}$ = 0.32 eV) also contributes to the total work function change. The effect of dipole Δϕ$_{Dip}$ is assumed to change the electron affinity χ (where the electron affinity is the energy difference between the vacuum level E$_{VAC}$ and the conduction band at the surface E$_{CBM}$). Thus the total work function change Δϕ due to the heterostructure formation is:

$$\Delta\phi = \Delta\chi + \Delta V_{BB} = \Delta\phi_{Dip} + \Delta V_{BB}$$

From this formula and the measured $\Delta\phi$ and $\Delta V_{BB}$ we calculate a surface dipole of $\Delta\phi_{Dip} = 0.2$ eV. This interface dipole is a consequence of interface electron redistribution between single layer of $MoS_2$ and GaN(0001). This does not imply any chemical bonding between the GaN and the $MoS_2$. Such charge redistribution at the interface between GaN and $MoS_2$ is also in agreement with GaSe/graphene[8,41] and graphene/$MoS_2$[28] heterostructures previously reported. It is interesting to underline also, that the Ga 3d core level peak shift by a lesser amount compared to the VBM after the heterostructure formation, this is probably related to this difference in the interface properties, *e.g.* this dipole formation.

## IV. CONCLUSIONS

In summary, the interaction of n-doped single layer $MoS_2$ on top of p-doped GaN layer was systematically studied via various characterization methods. An interfacial charge transfer was highlighted within the hybrid heterostructure using ARPES. Based on our measurements we propose a band diagram model to explain the charge doping effect deducing a conduction band discontinuity of about $\Delta E_C = 0.51$ eV in a type II alignment configuration. The experimental band alignment is determined by XPS/ARPES measurements comparing the effect of $MoS_2$ transfer on the electronic structure of GaN. Therefore, the heterointerface formation gives rise to an additional dipole change of 0.2 eV which could shift the band edges with respect to each other. The band alignment obtained in the present paper is essential information for building electronic and optoelectronic devices based on GaN/monolayer $MoS_2$ and to a larger extent, for understanding the electronic coupling in 2D/3D heterostructures.

## ACKNOWLEDGMENT

We acknowledge support from the Agence Nationale de la Recherche (ANR) under grants GANEX (Grant No. ANR-11-LABX-0014) and H2DH (ANR-15-CE24-0016), from the Region Ile-de-France in the framework of C'Nano IdF (nanoscience competence center of Paris Region), and from the European Union (FEDER 2007-2013). GANEX belongs to the public funded Investissements d'Avenir program managed by ANR. S.B is a member of Institut Universitaire de France (IUF).

**Figures captions:**

**Figure 1:** a) Schematic of our $MoS_2$/GaN heterostructure; b) Optical image of $MoS_2$ transferred on GaN; c) Raman spectrum of $MoS_2$/GaN; d) hyperspectral Raman map of the difference between the frequencies of the $A'_1$ and $E'$ mode features. e) Photoluminescence spectrum of $MoS_2$/GaN.

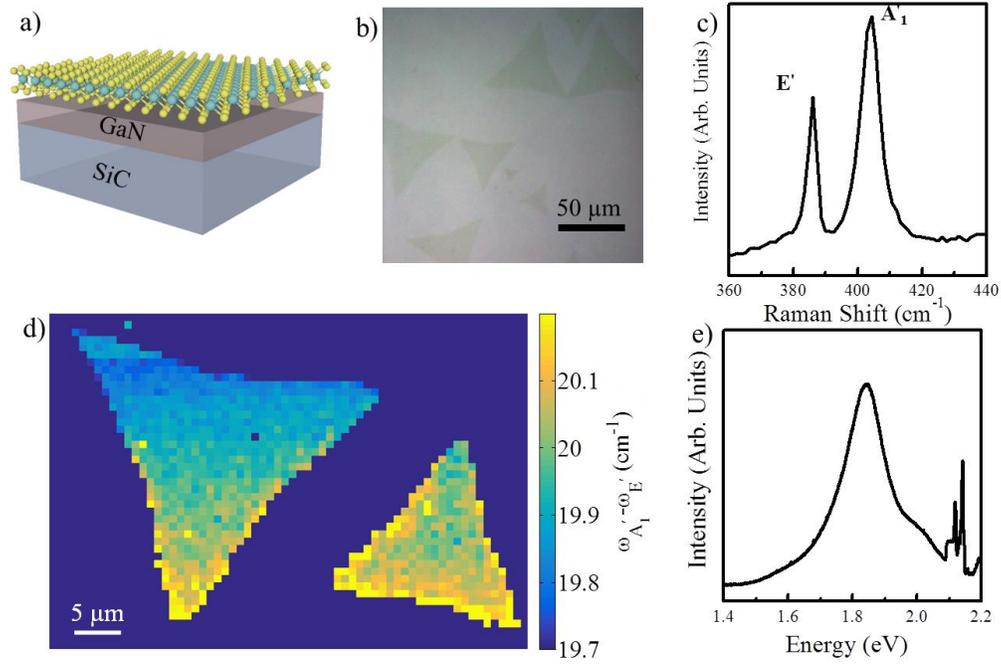

**Figure 2:** High-resolution XPS spectra of monolayered $MoS_2$/GaN heterostructure measured at hν =340 eV; a) Ga-3d for GaN and $MoS_2$/GaN, b) Mo-3d for $MoS_2$/GaN

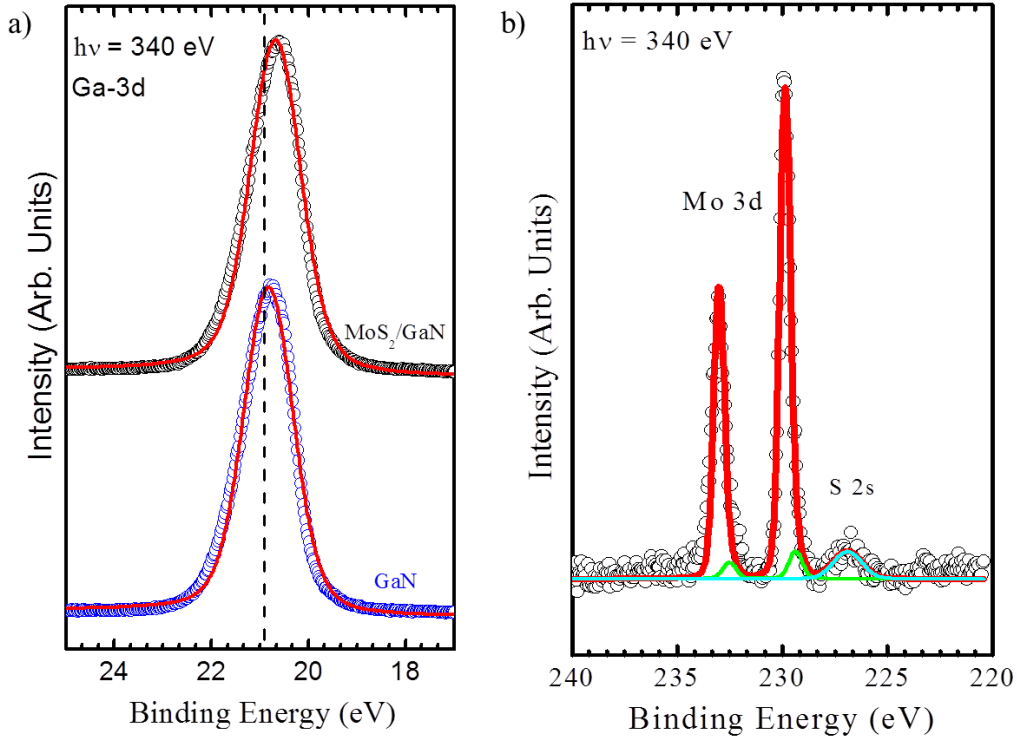

**Figure 3:** ARPES measurements of a) GaN and b) MoS$_2$/GaN measured at hv =50 eV; c) and d) Second-derivative spectra of a) and b) respectively, for better visibility of the bands.

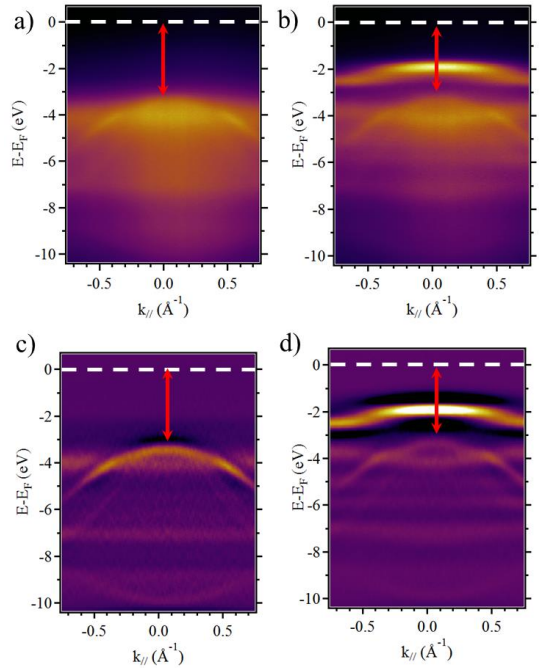

**Figure 4:** a) Integrated valence band at at k$_{//}$= 0 Å$^{-1}$ of MoS$_2$/GaN and GaN at hv =50 eV, b) Secondary electron cut-off *vs* kinetic energy of pristine GaN and MoS$_2$/GaN; c) Cathodo-luminescence of GaN.

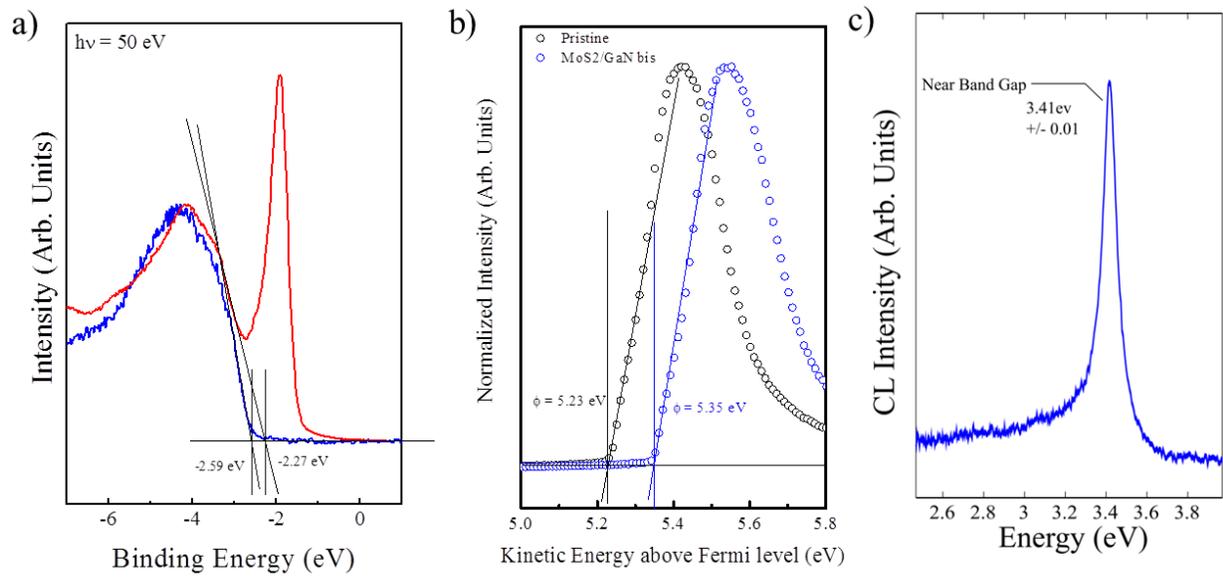

**Figure 5:** Schematic of band alignment diagram of MoS$_2$/GaN heterostructure obtained from XPS/ARPES and CL measurements. The band gap values of MoS$_2$ and GaN have been obtained considering their excitons binding energies[29,38]. The solid and dashed lines correspond respectively to the band bending after and before MoS$_2$ transfer highlighting a variation of the band bending in GaN after MoS$_2$ transfer.

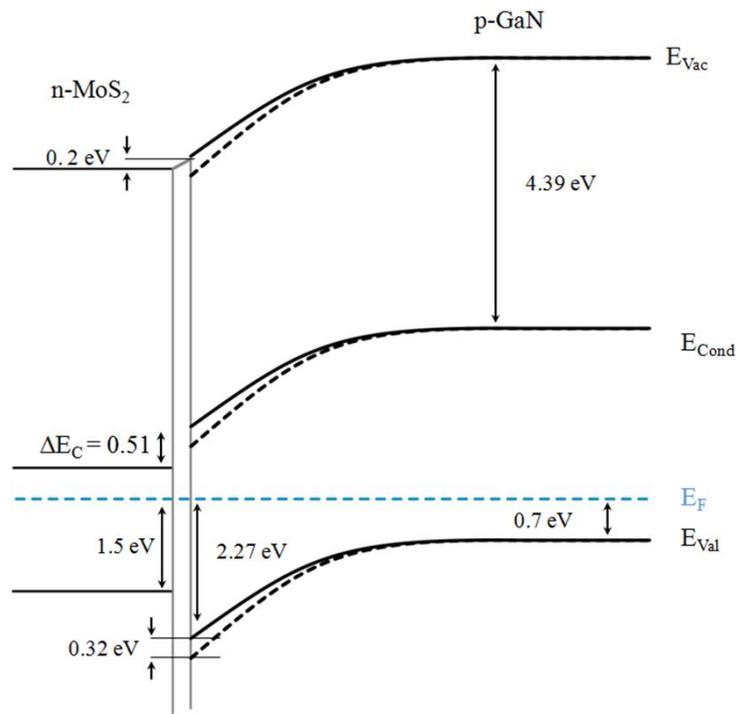

**References:**

(1) Bhimanapati, G. R.; Lin, Z.; Meunier, V.; Jung, Y.; Cha, J.; Das, S.; Xiao, D.; Son, Y.; Strano, M. S.; Cooper, V. R.; *et al.* Recent Advances in Two-Dimensional Materials beyond Graphene. *ACS Nano* **2015**, *9*, 11509–11539.

(2) Wang, Q. H.; Kalantar-Zadeh, K.; Kis, A.; Coleman, J. N.; Strano, M. S. Electronics and Optoelectronics of Two-Dimensional Transition Metal Dichalcogenides. *Nat. Nanotechnol.* **2012**, *7*, 699–712.

(3) Radisavljevic, B.; Radenovic, a; Brivio, J.; Giacometti, V.; Kis, a. Single-Layer MoS2 Transistors. *Nat. Nanotechnol.* **2011**, *6*, 147–150.

(4) Koppens, F. H.; Mueller, T.; Avouris, P.; Ferrari, a C.; Vitiello, M. S.; Polini, M. Photodetectors Based on Graphene, Other Two-Dimensional Materials and Hybrid Systems. *Nat Nanotechnol* **2014**, *9*, 780–793.

(5) Li, X.; Lin, M.-W.; Lin, J.; Huang, B.; Puretzky, A. A.; Ma, C.; Wang, K.; Zhou, W.; Pantelides, S. T.; Chi, M.; *et al.* Two-Dimensional GaSe/MoSe2 Misfit Bilayer Heterojunctions by van Der Waals Epitaxy. *Sci. Adv.* **2016**, *2*, e1501882.

(6) Wang, F.; Wang, Z.; Xu, K.; Wang, F.; Wang, Q.; Huang, Y.; Yin, L.; He, J. Tunable GaTe-MoS2 van Der Waals P-N Junctions with Novel Optoelectronic Performance. *Nano Lett.* **2015**, *15*, 7558–7566.

(7) Lee, C.-H.; Lee, G.; van der Zande, A. M.; Chen, W.; Li, Y.; Han, M.; Cui, X.; Arefe, G.; Nuckolls, C.; Heinz, T. F.; *et al.* Atomically Thin P–n Junctions with van Der Waals Heterointerfaces. *Nat. Nanotechnol.* **2014**, *9*, 676–681.

(8) Ben Aziza, Z.; Henck, H.; Pierucci, D.; Silly, M. G.; Lhuillier, E.; Patriarche, G.; Sirotti, F.; Eddrief, M.; Ouerghi, A. Van Der Waals Epitaxy of GaSe/Graphene Heterostructure: Electronic and Interfacial Properties. *ACS Nano* **2016**, *10*, 9679−9686.

(9) Henck, H.; Pierucci, D.; Chaste, J.; Naylor, C. H.; Avila, J.; Balan, A.; Silly, M. G.; Maria, C.; Sirotti, F.; Johnson, A. T. C.; *et al.* Electrolytic Phototransistor Based on Graphene-MoS2 van Der Waals P-N Heterojunction with Tunable Photoresponse P-N Heterojunction with Tunable Photoresponse. *Appl. Phys. Lett.* **2016**, *109*, 113103.

(10) Sarkar, D.; Xie, X.; Liu, W.; Cao, W.; Kang, J.; Gong, Y.; Kraemer, S.; Ajayan, P. M.; Banerjee, K. A Subthermionic Tunnel Field-Effect Transistor with an Atomically Thin Channel. *Nature* **2015**, *526*, 91–95.

(11) Ruzmetov, D.; Zhang, K.; Stan, G.; Kalanyan, B.; Bhimanapati, G. R.; Eichfeld, S. M.; Burke, R. A.; Shah, P. B.; O'Regan, T. P.; Crowne, F. J.; *et al.* Vertical 2D/3D Semiconductor Heterostructures Based on Epitaxial Molybdenum Disulfide and Gallium Nitride. *ACS Nano* **2016**, *10*, 3580–3588.

(12) Lee, C. H.; Krishnamoorthy, S.; O'Hara, D. J.; Johnson, J. M.; Jamison, J.; Myers, R. C.; Kawakami, R. K.; Hwang, J.; Rajan, S. Molecular Beam Epitaxy of 2D-Layered Gallium Selenide on GaN Substrates. *arXiv:1610.06265* **2016**.

(13) Lee, E. W.; Lee, C. H.; Paul, P. K.; Ma, L.; McCulloch, W. D.; Krishnamoorthy, S.; Wu, Y.; Arehart, A. R.; Rajan, S. Layer-Transferred MoS2/GaN PN Diodes. *Appl. Phys. Lett.* **2015**, *107*, 103505.

(14) Tangi, M.; Mishra, P.; Tseng, C.-C.; Ng, T. K.; Hedhili, M. N.; Anjum, D. H.; Alias, M. S.; Wei, N.; Li, L.-J.; Ooi, B. S. Band Alignment at GaN/Single-Layer WSe$_2$ Interface. *ACS Appl. Mater. Interfaces* **2017**,

acsami.6b15370.

(15) Tangi, M.; Mishra, P.; Ng, T. K.; Hedhili, M. N.; Janjua, B.; Alias, M. S.; Anjum, D. H.; Tseng, C. C.; Shi, Y.; Joyce, H. J.; *et al.* Determination of Band Offsets at GaN/single-Layer MoS2 Heterojunction. *Appl. Phys. Lett.* **2016**, *109*.

(16) Jariwala, D.; Marks, T. J.; Hersam, M. C. Mixed-Dimensional van Der Waals Heterostructures. *Nat. Mater.* **2016**, *16*, 170–181.

(17) Kang, M. S.; Lee, C. H.; Park, J. B.; Yoo, H.; Yi, G. C. Gallium Nitride Nanostructures for Light-Emitting Diode Applications. *Nano Energy* **2012**, *1*, 391–400.

(18) Pierucci, D.; Henck, H.; Naylor, C. H.; Sediri, H.; Lhuillier, E.; Balan, A.; Rault, J. E.; Dappe, Y. J.; Bertran, F.; Le Févre, P.; *et al.* Large Area Molybdenum Disulphide-Epitaxial Graphene Vertical Van Der Waals Heterostructures. *Sci. Rep.* **2016**, *6*, 26656.

(19) Hong, X.; Kim, J.; Shi, S.-F.; Zhang, Y.; Jin, C.; Sun, Y.; Tongay, S.; Wu, J.; Zhang, Y.; Wang, F. Ultrafast Charge Transfer in Atomically Thin MoS2/WS2 Heterostructures. *Nat. Nanotechnol.* **2014**, *9*, 1–5.

(20) Georgiou, T.; Jalil, R.; Belle, B. D.; Britnell, L.; Gorbachev, R. V; Morozov, S. V; Kim, Y.-J.; Gholinia, A.; Haigh, S. J.; Makarovsky, O.; *et al.* Vertical Field-Effect Transistor Based on Graphene-WS$_2$ Heterostructures for Flexible and Transparent Electronics. *Nat. Nanotechnol.* **2012**, *8*, 100–103.

(21) Chiu, M.-H.; Zhang, C.; Shiu, H.-W.; Chuu, C.-P.; Chen, C.-H.; Chang, C.-Y. S.; Chen, C.-H.; Chou, M.-Y.; Shih, C.-K.; Li, L.-J. Determination of Band Alignment in the Single-Layer MoS2/WSe2 Heterojunction. *Nat. Commun.* **2015**, *6*, 7666.

(22) Lieten, R. R.; Motsnyi, V.; Zhang, L.; Cheng, K.; Leys, M.; Degroote, S.; Buchowicz, G.; Dubon, O.; Borghs, G. Mg Doping of GaN by Molecular Beam Epitaxy. *J. Phys. D. Appl. Phys.* **2011**, *44*, 135406.

(23) Han, G. H.; Kybert, N. J.; Naylor, C. H.; Lee, B. S.; Ping, J.; Park, J. H.; Kang, J.; Lee, S. Y.; Lee, Y. H.; Agarwal, R.; *et al.* Seeded Growth of Highly Crystalline Molybdenum Disulphide Monolayers at Controlled Locations. *Nat. Commun.* **2015**, *6*, 6128.

(24) Wang, Y.; Cong, C.; Qiu, C.; Yu, T. Raman Spectroscopy Study of Lattice Vibration and Crystallographic Orientation of Monolayer mos2 under Uniaxial Strain. *Small* **2013**, *9*, 2857–2861.

(25) Lee, C.; Yan, H.; Brus, L. E.; Heinz, T. F.; Hone, K. J.; Ryu, S. Anomalous Lattice Vibrations of Single-and Few-Layer MoS2. *ACS Nano* **2010**, *4*, 2695–2700.

(26) Ben Aziza, Z.; Henck, H.; Di, D.; Pierucci, D.; Chaste, J.; Naylor, C. H.; Balan, A.; Dappe, Y. J.; Johnson, A. T. C.; Ouerghi, A. Bandgap Inhomogeneity of MoS 2 Monolayer on Epitaxial Graphene Bilayer in van Der Waals P-N Junction. *Carbon N. Y.* **2016**, *110*, 396–403.

(27) Eda, G.; Yamaguchi, H.; Voiry, D.; Fujita, T.; Chen, M.; Chhowalla, M. Photoluminescence from Chemically Exfoliated MoS 2. *Nano Lett.* **2011**, *11*, 5111–5116.

(28) Pierucci, D.; Henck, H.; Avila, J.; Balan, A.; Naylor, C. H.; Patriarche, G.; Dappe, Y. J.; Silly, M. G.; Sirotti, F.; Johnson, A. T. C.; *et al.* Band Alignment and Minigaps in Monolayer MoS2-Graphene van Der Waals Heterostructures. *Nano Lett.* **2016**, *16*, 4054–4061.

(29) Kushvaha, S. S.; Kumar, M. S.; Shukla, A. K.; Yadav, B. S.; Singh, D. K.; Jewariya, M.; Ragam, S. R.;

Maurya, K. K. Structural, Optical and Electronic Properties of Homoepitaxial GaN Nanowalls Grown on GaN Template by Laser Molecular Beam Epitaxy. *RSC Adv.* **2015**, *5*, 87818–87830.

(30) Mishra, M.; Krishna, T. C. S.; Aggarwal, N.; Gupta, G. Surface Chemistry and Electronic Structure of Nonpolar and Polar GaN Films. *Appl. Surf. Sci.* **2015**, *345*, 440–447.

(31) Mattila, S.; Leiro, J. a.; Heinonen, M.; Laiho, T. Core Level Spectroscopy of MoS2. *Surf. Sci.* **2006**, *600*, 5168–5175.

(32) Kim, I. S.; Sangwan, V. K.; Jariwala, D.; Wood, J. D.; Park, S.; Chen, K.; Shi, F.; Ruiz-zepeda, F.; Ponce, A.; Jose-, M.; *et al.* Influence of Stoichiometry on the Optical and Electrical Properties of Chemical Vapor Deposition Derived MoS 2. *ACS Nano* **2014**, *8*, 10551–10558.

(33) Zhou, W.; Zou, X.; Najmaei, S.; Liu, Z.; Shi, Y.; Kong, J.; Lou, J. Intrinsic Structural Defects in Monolayer Molybdenum Disul Fi de. *Nano Lett.* **2013**, *13*, 2615–2622.

(34) Addou, R.; Mcdonnell, S.; Barrera, D.; Guo, Z.; Azcatl, A.; Wang, J.; Zhu, H.; Hinkle, C. L.; Quevedo-lopez, M.; Alshareef, H. N.; *et al.* Impurities and Electronic Property Variations of Natural MoS 2 Crystal Surfaces. *ACS Nano* **2015**, 9124–9133.

(35) Levasseur, a; Vinatier, P.; Gonbeau, D. X-Ray Photoelectron Spectroscopy: A Powerful Tool for a Better Characterization of Thin Film Materials. *Bull. Mater. Sci.* **1999**, *22*, 607–614.

(36) Baker, M. A.; Gilmore, R.; Lenardi, C.; Gissler, W. XPS Investigation of Preferential Sputtering of S from MoS 2 and Determination of MoS X Stoichiometry from Mo and S Peak Positions. *Appl. Surf. Sci.* **1999**, *150*, 255–262.

(37) Fleischauer, P. D.; Lince, J. R. A Comparison of Oxidation and Oxygen Substitution in MoS 2 Solid Film Lubricants. *Tribol. Int.* **1999**, *32*, 627–636.

(38) Ugeda, M. M.; Bradley, A. J.; Shi, S.; Jornada, F. H.; Zhang, Y.; Qiu, D. Y.; Ruan, W.; Mo, S.; Hussain, Z.; Shen, Z.; *et al.* Giant Bandgap Renormalization and Excitonic E Ects in a Monolayer Transition Metal Dichalcogenide Semiconductor. *Nat. Mater.* **2014**, *13*, 1091–1095.

(39) Strite, S.; Morkoç, H. GaN, AlN, and InN: A Review. *J. Vac. Sci. Technol. B Microelectron. Nanom. Struct. Process. Meas. Phenom.* **1992**, *10*, 1237–1266.

(40) Kummell, T.; Hutten, U.; Heyer, F.; Derr, K.; Neubieser, R. M.; Quitsch, W.; Bacher, G. Carrier Transfer across a 2D-3D Semiconductor Heterointerface: The Role of Momentum Mismatch. *Phys. Rev. B - Condens. Matter Mater. Phys.* **2017**, *95*, 1–5.

(41) Si, C.; Lin, Z.; Zhou, J.; Sun, Z. Controllable Schottky Barrier in GaSe/graphene Heterostructure: The Role of Interface Dipole. *2D Mater.* **2016**, *4*, 15027.

# Interface Dipole and Band Bending in Hybrid p-n Heterojunction MoS$_2$/GaN(0001)

Hugo Henck[1], Zeineb Ben Aziza[1], Olivia Zill[2], Debora Pierucci[3], Carl H. Naylor[4], Mathieu G. Silly[5], Noelle Gogneau[1], Fabrice Oehler[1], Stephane Collin[1], Julien Brault[6], Fausto Sirotti[5], François Bertan[5], Patrick Lefèvre[5], Stéphane Berciaud[2], A.T Charlie Johnson[4], Emmanuel Lhuillier[7], Julien E. Rault[5] and Abdelkarim Ouerghi[1]

[1]Centre de Nanosciences et de Nanotechnologies, CNRS, Univ. Paris-Sud, Université Paris-Saclay, C2N – Marcoussis, 91460 Marcoussis, France
[2]Université de Strasbourg, CNRS, Institut de Physique et Chimie des Matériaux de Strasbourg (IPCMS), UMR 7504, F-67000 Strasbourg, France
[3]CELLS - ALBA Synchrotron Radiation Facility, Carrer de la Llum 2-26, 08290 Cerdanyola del Valles, Barcelona, Spain
[4]Department of Physics and Astronomy, University of Pennsylvania, 209S 33rd Street, Philadelphia, Pennsylvania 19104, USA
[5] Synchrotron-SOLEIL, Saint-Aubin, BP48, F91192 Gif sur Yvette Cedex, France
[6]Université Côte d'Azur, CNRS, CRHEA, 06560 Valbonne Sophia Antipolis, France
[7]Sorbonne Universités, UPMC Univ. Paris 06, CNRS-UMR 7588, Institut des NanoSciences de Paris, 4 place Jussieu, 75005 Paris, France

*Corresponding author, E-mail: abdelkarim.ouerghi@c2n.upsaclay.fr , fax number: +33169636006

**X-ray photoemission spectroscopy (XPS)**

X-ray photoemission spectroscopy (XPS) has been performed at hν = 1300 eV on p-doped GaN(0001), in synchrotron radiation facility at SOLEIL. The detected photoelectron intensity in this energy configuration includes the contribution of electrons coming from several of nanometers in the GaN layers. The valence band maximum of the bulk GaN band structure is then obtained by fitting the leading edge of the integrated photoelectron intensity spectra at -0.7 eV with respect to the Fermi level.

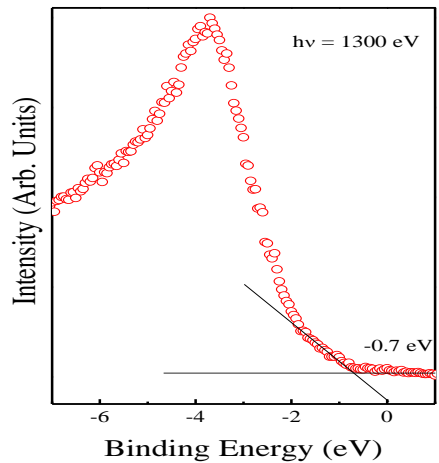

**Figure S1:** Integrated photoelectron intensity measured at hv =1300 eV to determine the Valence band maximum of bulk GaN.

**Cathodoluminescence**

The low acceleration voltage (2 kV) allows probing the topmost layers of the GaN surface as illustrated in Figure S2. Monte Carlo simulations (see methods) indicates that 50% of the CL signal originates from the first 10 nm of GaN (S2(a)). The panchromatic CL image (S2(b)) centered on the GaN signal reveals the clear shape of the $MoS_2$ flake with a brighter intensity. The average spectra extracted from the bare GaN and the GaN/$MoS_2$ area (blue and red in S2(c)) confirm the GaN bandgap value of 3.41 eV with a 2.5 time enhancement of the GaN peak intensity below the $MoS_2$ flake compared to that of the bare GaN surface.

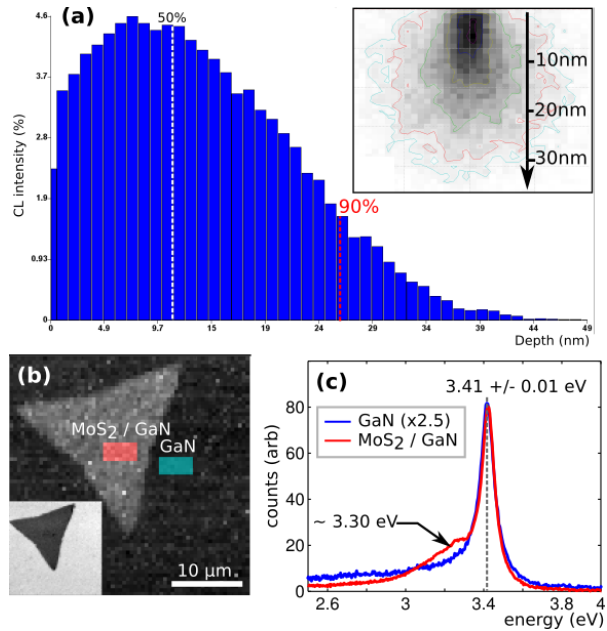

**Figure S2:** (a) Monte Carlo simulation of the origin of the CL signal from a GaN layer at 2 kV acceleration voltage. The histogram represents the fraction of CL signal originating from a given slice along the normal direction, 50% (90%) of the total signal is emitted from the first 10 (25) nm. The inset shows the typical in-plane extension (dark corresponds to maximum CL intensity). (b) Panchromatic CL map (2.6-4.0 eV) of a $MoS_2$ flake on GaN. The inset shows the corresponding SEM image. (c) Room temperature CL spectra of GaN and GaN/$MoS_2$ obtained from the highlighted zones of the panchromatic image shown in (b).